%% file: main.tex
\documentclass[runningheads]{llncs}
\input{packages}

\begin{document}
\title{SISA: A Scale-In Systolic Array for GEMM Acceleration}
%
%
\author{Luigi Altamura\orcidID{0009-0000-2070-3932} \and
Alessio Cicero\orcidID{0009-0002-9176-0605} \and
Mateo Vázquez Maceiras\orcidID{0000-0002-5681-9344}\and
Mohammad Ali Maleki\orcidID{0000-0002-9019-3605}\and
Pedro Trancoso\orcidID{0000-0002-2776-9253}}
\authorrunning{Altamura et al.}
%
\institute{Department of Computer Science and Engineering,\\Chalmers University of Technology and University of Gothenburg\\ Gothenburg, Sweden\\
\email{\{altamura,alessio.cicero,maceiras,mohammad.ali.maleki,ppedro\}@chalmers.se}}
\maketitle              
\begin{abstract}    
\input{sections/00_abstract}

\end{abstract}

\input{sections/01_introduction}
\input{sections/02_background}

\input{sections/03_sisa}

\input{sections/04_results}
\input{sections/05_conclusion}
\input{sections/06_acknowledgments}
%
%

%
\bibliographystyle{splncs04}
\bibliography{bibliography}

\end{document}

%% file: packages.tex
\usepackage[dvipsnames]{xcolor}
\usepackage{acronym}

\input{acronyms}
\usepackage[T1]{fontenc}
\usepackage{makecell}
\usepackage{graphicx}
\usepackage{subcaption}
\usepackage{algorithm}
\usepackage{algorithmic}
\usepackage{booktabs}
\usepackage{amssymb}
\usepackage{amsmath} 

\newif\ifvlms
\vlmsfalse

%% file: acronyms.tex
\newacro{HPC}[HPC]{High Performance Computing}

\newacro{EDP}[EDP]{Energy-Delay Product}

\newacro{AI}[AI]{Artificial Intelligence}
\newacro{ML}[ML]{Machine Learning}
\newacro{DL}[DL]{Deep Learning}

\newacro{MLP}[MLP]{Multi-Layer Perceptron}
\newacro{MHA}[MHA]{Multi-Head Attention}
\newacro{MSA}[MSA]{Multi-Head Self-Attention}

\newacro{FFN}[FFN]{Feed-Forward Network}
\acrodefplural{FFN}[FFNs]{Feed-Forward Networks}
\newacro{CNN}[CNN]{Convolutional Neural Network}
\acrodefplural{CNN}[CNNs]{Convolutional Neural Networks}

\newacro{LLM}[LLM]{Large Language Model}
\acrodefplural{LLM}[LLMs]{Large Language Models}
\newacro{VPU}[VPU]{Vector Processing Unit}
\acrodefplural{VPU}[VPUs]{Vector Processing Units}
\newacro{VLM}[VLM]{Vision Language Model}
\acrodefplural{VLM}[VLMs]{Vision Language Models}
\newacro{ViT}[ViT]{Vision Transformer}
\acrodefplural{ViT}[ViTs]{Vision Transformers}

\newacro{GEMM}[GEMM]{General Matrix-Matrix Multiplication}
\acrodefplural{GEMM}[GEMMs]{General Matrix-Matrix Multiplications}
\newacro{GEMV}[GEMV]{General Matrix-Vector Multiplication}
\acrodefplural{GEMV}[GEMVs]{General Matrix-Vector Multiplications}
\newacro{SA}[SA]{Systolic Array}
\acrodefplural{SA}[SAs]{Systolic Arrays}
\newacro{SOSA}[SOSA]{Scale-Out Systolic Array}
\acrodefplural{SOSA}[SOSAs]{Scale-Out Systolic Arrays}
\newacro{SISA}[SISA]{Scale-In Systolic Array}

\newacro{OS}[OS]{Output Stationary}
\newacro{WS}[WS]{Weight Stationary}

\newacro{TPU}[TPU]{Tensor Processing Unit}
\acrodefplural{TPU}[TPUs]{Tensor Processing Units}

\newacro{PE}[PE]{Processing Element}
\acrodefplural{PE}[PEs]{Processing Elements}
\newacro{MAC}[MAC]{Multiply-Accumulate}
\newacro{FPU}[FPU]{Floating-Point Unit}
\acrodefplural{FPU}[FPUs]{Floating-Point Units}

\newacro{TTFT}[TTFT]{Time To First Token}

%% file: sections/00_abstract.tex
The currently dominant AI/ML workloads, such as \acp{LLM}, rely on the efficient execution of \ac{GEMM} operations. Thus, most systems are equipped with dedicated matrix hardware accelerators based on square \acp{SA} of \acp{PE}. 
While this organization was effective for traditional Deep Neural Networks (DNNs), \acp{LLM} introduce input-dependent and highly skewed matrices, leading to underutilized \ac{SA} resources.

To address this challenge, we propose SISA (Scale-In Systolic Array), a novel \ac{SA} architecture that partitions the traditional square array into horizontal rectangular slabs. With minimal overhead, SISA exposes parallelism through independently scheduled slabs for efficient execution of small or skewed matrix shapes, while retaining full-array operation for large \acp{GEMM}. SISA achieves up to 8.52$\times$ speedup and 93\% energy-delay-product (EDP) reduction for representative \acp{LLM} compared to a state-of-the-art monolithic SA with the same number of \acp{PE}.


\keywords{systolic array, GEMM, LLM, accelerator, adaptive architecture}

%% file: sections/01_introduction.tex
\section{Introduction}
\label{sec:intro}

In recent years, the use of AI-based applications has increased rapidly. 

\acfp{LLM}, which form the backbone of widely deployed interactive systems such as ChatGPT, are quite demanding on the computing systems.

\ac{LLM} inference consists of two distinct phases: \emph{prefill} and \emph{decode}, each with fundamentally different characteristics. During prefill, the model processes the entire input prompt using dense \acfp{GEMM}, which are compute-bound. In contrast, decode generates tokens sequentially using \acfp{GEMV}, making it predominantly memory-bandwidth bound. In addition to \acp{GEMV}, the decode also repeatedly accesses the key–value (KV) cache, which increases in size at each decode step.

These characteristics naturally map the phases to different architectural primitives. Prefill benefits from \acf{SA}-based accelerators that maximize throughput for large \acp{GEMM}, whereas decode is better suited to \acp{VPU} that efficiently handle \acp{GEMV}.

However, for \ac{LLM} systems supporting multiple concurrent requests, batching may be exploited to increase arithmetic intensity and amortize memory traffic~\cite{sarathi_serve,deep_speed_inference}. 
This leads to having \ac{GEMM} operations in the decode phase as well, making a \ac{SA}-based accelerator the primary architecture for \acp{LLM}.
As a result, many modern AI accelerators incorporate \acp{SA}. The most well-known example is the Google TPU, which integrates a large square \ac{SA} (e.g., 128×128 and 256×256 \acp{PE})~\cite{tpuv4} that is highly efficient for executing the large \acp{GEMM} common in \ac{LLM} workloads.

\begin{figure*}[t]
    \centering
    \begin{subfigure}{0.45\textwidth}
        \centering
        \includegraphics[width=\linewidth]{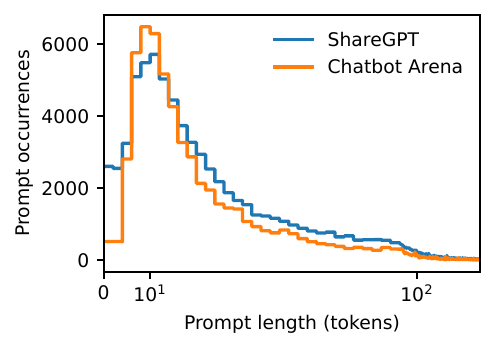}
        \caption{}
    \end{subfigure}\hspace{0.05\textwidth}
    \begin{subfigure}{0.45\textwidth}
        \centering
        \includegraphics[width=\linewidth]{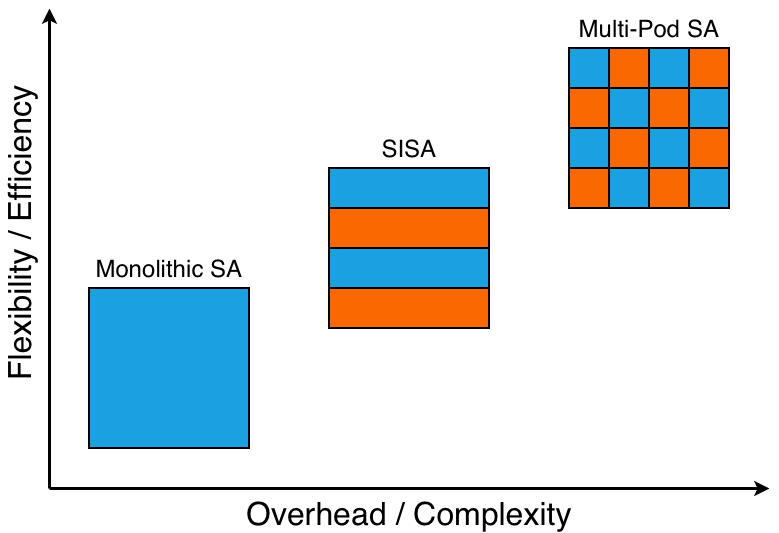}
        \caption{}
    \end{subfigure}\hspace{0.04\textwidth}
    \caption{(a) Prompt length derived from interactive chatbot requests and (b) alternative SA designs.}
    \label{fig:motivation}
    \vspace{-12pt}
\end{figure*}

Nevertheless, when matrices are highly-skewed or smaller than the \ac{SA} size, a significant fraction of \acp{PE} remain idle. Previous works have shown that such \ac{SA}-matrix shape mismatches lead to considerable under-utilization~\cite{mutlu_tpu_utilization}, and

these mismatches are common in practice.
Figure~\ref{fig:motivation}a shows that real-world chatbot prompts are typically short (mean 41.8, median 12 tokens)~\cite{wang2025llmservingoptimizationvariable}, leading to skewed \ac{GEMM}s during the \ac{LLM} prefill phase.

Consequently, the \ac{LLM} prefill phase often involves skewed matrices (e.g., $(12,8192)$ for a 12-token prompt in Llama3.2-3B).
In addition, the number of tokens per prompt varies across requests. When multi-batching is not possible or when latency-sensitive prefill is required, matrix shapes remain highly skewed, since increasing batching may degrade Quality-of-Service by increasing the \ac{TTFT}.

When decode requests are batched to amortize memory access costs in a \ac{GEMM}-based \ac{LLM} decode phase, the resulting matrix shapes depend on the batch size, with small batches yielding highly skewed matrices.
However, two factors prevent aggressively large batching (e.g., decode batch sizes beyond 32–64). First, larger batch sizes increase response time, so batching is kept moderate to balance QoS and efficiency. Second, the decode phase accesses the KV cache, which can become large depending on the request. Since each request maintains its own KV cache, batching requires multiple such structures to be stored simultaneously in memory, further limiting the batch size.

In order to address these and other mismatch issues between \acp{SA} and \ac{GEMM} shapes of matrices, researchers have proposed different architectures that can adapt the arrangement of available \acp{PE} to better match the shapes of the matrices~\cite{savector,Planaria,ReDAs,DyNNamic,FlexSA,SARA,halo,sosa}. In Figure~\ref{fig:motivation}b we show the tradeoff between flexibility/efficiency and overhead/complexity. Monolithic \ac{SA} accelerators serves as the baseline, whereas flexible accelerators such as the multi-pod \ac{SA} offers more flexibility at the cost of increased buffer and interconnection network overheads.

To address this challenge, we propose SISA, a new \ac{SA} architecture that extends the traditional $n \times n$ square design with a \textit{scale-in} mechanism that partitions the array into horizontal smaller rectangular sub-arrays, referred to
as slabs, capable of parallel, independently scheduled execution. This slab-level parallelism enables SISA to efficiently process highly skewed matrix shapes by increasing effective parallelism relative to a monolithic array. When full-array execution is advantageous, SISA seamlessly operates as a conventional systolic array, preserving peak GEMM performance while incurring only minimal additional on-chip buffering overhead. This design point is represented in Figure~\ref{fig:motivation}b.

The main contributions of this work are:
\begin{itemize}
\item A novel \ac{SA} architecture that operates either as a unified array or partitioned into slabs that better adapt to skewed matrix shapes.
\item A scheduling and tiling mechanism that exploits parallelism across slabs.
\item An instance of SISA with $128 \times $128 \acp{PE} and its  evaluation using state-of-the-art (SOTA) LLM models.
\end{itemize}

We have evaluated SISA with \ac{GEMM} operations from \acp{LLM}, and observed a speedup of up to $8.52\times$ and a reduction in energy-delay-product (EDP) up to 93\% when compared to a TPU with the same amount of compute and memory resources. In the worst case, SISA achieves comparable performance to the TPU with only an 8.47\% higher EDP. When compared to alternative SOTA flexible \ac{SA} architectures, SISA performs better in most cases, achieving speedups up to $2.61\times$ while also exhibiting lower area overhead.

%% file: sections/02_background.tex
\section{Related Work}
\label{sec:related}

\input{tables/comparison}

To meet the computational requirements of \acp{LLM}, \acp{SA} appear as a fitting architecture to build ML accelerators.
\acp{SA} are architectures composed by a set of interconnected \acp{PE}.
Thanks to their regular structure and communication patterns, \acp{SA} can offer high computational capabilities for \ac{GEMM} while keeping energy consumption efficient compared to other architectures: in \acp{SA}, results are fed directly to the next computational unit without intermediate storage, following a systolic wavefront~\cite{kung_why_sa}.

However, while monolithic \acp{SA} are a matching architecture for computing \ac{GEMM}, multiple variations have been presented in recent works.
These variations of \acp{SA} aim to better match the characteristics and requirements of different workloads like \acp{LLM}.
In this section, we present the most relevant proposals, which are summarized in Table~\ref{tab:comparison}.


\subsubsection{Monolithic \aclp{SA}.}
\label{subsubsec:monolithic}

Early DL accelerators adopted large, fixed-size \acp{SA} to maximize GEMM throughput, like Google's TPU~\cite{tpuv4} and Gemmini~\cite{gemmini}.
Scaling up a single array provides high compute density and high throughput for large workloads, but it suffers from substantial under-utilization for small or irregular matrices, where many \acp{PE} remain idle due to poor alignment between workload shape and array dimensions~\cite{mutlu_tpu_utilization}.

\subsubsection{Multi-Pod Systolic Architectures.}
\label{subsubsec:multipod}

To improve utilization, one approach is to build architectures based on smaller \acp{SA}, named pods, and distribute computation across them.
Example of this approach are Scale-Out Systolic Arrays (SOSA)~\cite{sosa} and Vectored Systolic Arrays (SAVector)~\cite{savector}.
Overall, multi-pod systolic architectures enhance parallel tile-level processing but inherently increase data movement and off-chip memory access due to duplicated tile loads across pods.
In addition, their fixed pod granularity limits utilization on large \acp{GEMM} where a unified array would provide higher data reuse.
These architectures increase flexibility but introduce redundant data movement and elevated memory traffic, resulting in a reduction of the overall efficiency.

\subsubsection{Reconfigurable \aclp{SA}.}
\label{subsubsec:reconfigurable}

To be able to provide efficient execution of small, medium, and large \acp{GEMM}, reconfigurable \acp{SA} support runtime reconfiguration of large \acp{SA} into smaller tiles, increasing utilization while retaining the performance of monolithic \acp{SA} for large \acp{GEMM}.
However, this reconfigurability comes at a cost, which depends on the specific approach followed. Planaria~\cite{Planaria}, which targets multi-workload execution, is limited to coarse-grained reconfiguration, i.e., not flexible enough, and thus still suffers from under-utilization issues. Works like FlexSA~\cite{FlexSA}, HALO~\cite{halo}, DyNNamic~\cite{DyNNamic}, and SARA~\cite{SARA} support finer reconfigurability, but that comes with an increased area overhead. ReDas aims to solve this problem to some extend, but does it so at the cost of not being able to use all \acp{PE} in multiple configurations~\cite{ReDAs}.

\subsubsection{This Work.}
\label{sec:design_considerations}

Contrary to existing works, SISA provides a \ac{SA} architecture that (1) exhibits the compute density of a large \ac{SA} for big \ac{GEMM} operations, (2) enables more flexible execution modes that mitigate under-utilization, and (3) does so with low area overhead.

%% file: tables/comparison.tex
\setlength{\tabcolsep}{6pt}
\begin{table}[t]
\centering
\small
\caption{Comparison of SAs across key design requirements. For relative requirements, like the area overhead, we take as reference an equivalent monolithic SA.}
\label{tab:comparison}
\begin{tabular}{lccc}
\toprule
Architecture &
\shortstack{Fine-Grained Array/Pods\\Reconfigurability} &
\shortstack{Low Area\\Overhead} &
\shortstack{Low Data\\Movement} \\
\midrule
TPU~\cite{tpuv4}   & \textcolor{BrickRed}{\sffamily X}  & 
\textcolor{OliveGreen}{\checkmark} & \textcolor{OliveGreen}{\checkmark} \\
Gemmini~\cite{gemmini}   & \textcolor{BrickRed}{\sffamily X}  & 
\textcolor{OliveGreen}{\checkmark} & \textcolor{OliveGreen}{\checkmark} \\
\midrule
SOSA~\cite{sosa}         & \textcolor{OliveGreen}{\checkmark} & 
\textcolor{BrickRed}{\sffamily X}  & \textcolor{BrickRed}{\sffamily X}  \\
SAVector~\cite{savector} & \textcolor{OliveGreen}{\checkmark} & 
\textcolor{OliveGreen}{\checkmark}  & \textcolor{BrickRed}{\sffamily X}  \\
\midrule
Planaria~\cite{Planaria} & \textcolor{BrickRed}{\sffamily X}  & 
\textcolor{OliveGreen}{\checkmark} & \textcolor{OliveGreen}{\checkmark} \\
FlexSA~\cite{FlexSA}     & \textcolor{OliveGreen}{\checkmark} & 
\textcolor{BrickRed}{\sffamily X}  & \textcolor{OliveGreen}{\checkmark} \\
HALO~\cite{halo}         & \textcolor{OliveGreen}{\checkmark} & 
\textcolor{BrickRed}{\sffamily X}  & \textcolor{OliveGreen}{\checkmark} \\
DyNNamic~\cite{DyNNamic} & \textcolor{OliveGreen}{\checkmark} & 
\textcolor{BrickRed}{\sffamily X}  & \textcolor{OliveGreen}{\checkmark} \\
SARA~\cite{SARA}         & \textcolor{OliveGreen}{\checkmark} & 
\textcolor{BrickRed}{\sffamily X}  & \textcolor{OliveGreen}{\checkmark} \\
ReDas~\cite{ReDAs}       & \textcolor{OliveGreen}{\checkmark} & 
\textcolor{BrickRed}{\sffamily X} & \textcolor{OliveGreen}{\checkmark} \\
\midrule
\textbf{SISA (this work)} & \textcolor{OliveGreen}{\checkmark} & 
\textcolor{OliveGreen}{\checkmark} & \textcolor{OliveGreen}{\checkmark} \\
\bottomrule
\end{tabular}
\end{table}

%% file: sections/03_sisa.tex
\begin{figure*}[t]
    \centering
    \includegraphics[width=1\textwidth]{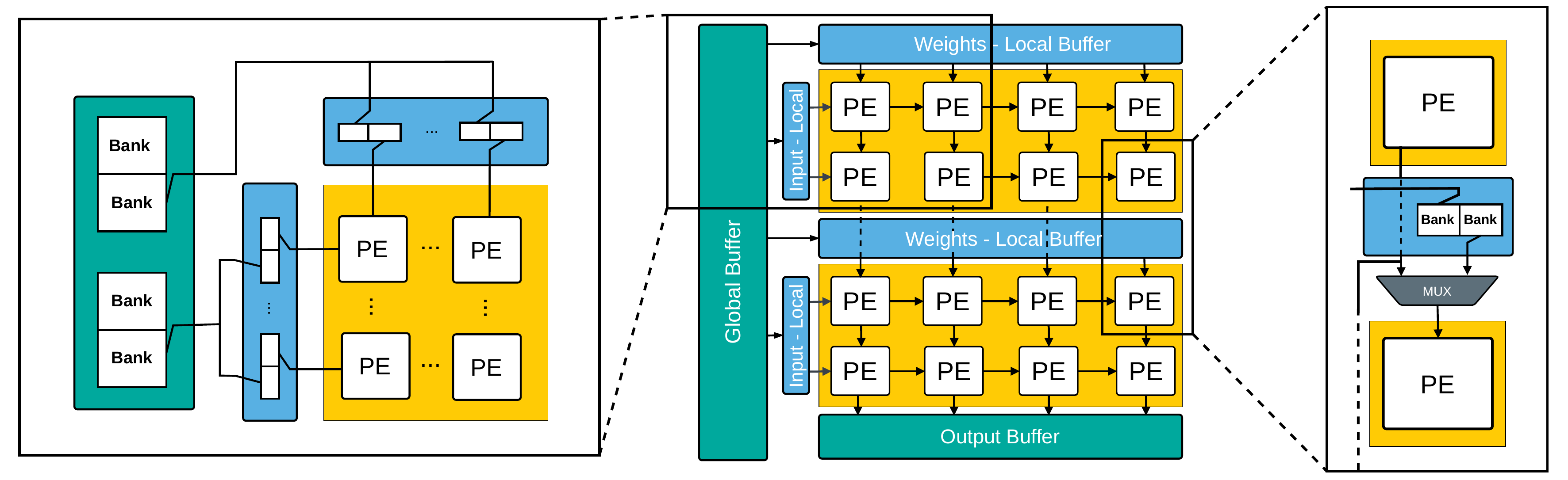}
    \caption{Overview of the SISA architecture with detailed views of the memory hierarchy and slab fusion mechanism.}
    \label{fig:arch_overview}
    \vspace{-12pt}
\end{figure*}

\section{Scale-In Systolic Array}
\label{sec:sisa}

\subsection{Architecture}

At a high level, SISA consists of a fixed pool of \acp{PE} organized as a logical two-dimensional \ac{SA} and horizontally partitioned into rectangular slabs to efficiently process skewed matrices. Each slab forms an independently controllable systolic unit with local buffers for activations and weights. Through multiplexing and power-gating, slabs can be fused into larger logical arrays or operated independently to exploit parallelism within a \ac{GEMM}. Unused slabs can be power-gated to reduce idle energy. SISA operates as a CPU-hosted accelerator with access to external memory. Figure~\ref{fig:arch_overview} provides an overview of the architecture.

\subsubsection{Slab Structure and Dataflow.}
Each slab has a rectangular organization, spanning the full width of the \ac{SA} while occupying a fraction of its height since \ac{LLM} workloads are typically narrow in one dimension but wide in the other. As described earlier, each slab includes local buffers for activations and weights, while output values are written directly to the global output buffer.
SISA employs an Output-Stationary (OS) dataflow. Under this dataflow, each \ac{PE} is responsible for a unique output element and maintains the corresponding accumulator locally for the duration of the \ac{GEMM} processing. As a result, partial sums do not traverse inter-PE interconnects; only activations and weights are streamed through the array. This property simplifies slab composition and enables modular and scalable execution across fused and independent slabs. The OS dataflow is particularly effective when the \ac{SA} matches, even partially, the output tensor dimensions of the \ac{GEMM}. 

\subsubsection{Slab Fusion Mechanism.}

Slabs can be fused along the height dimension of the \ac{SA} to form larger logical slabs. As shown in Figure~\ref{fig:arch_overview}, this is enabled through a simple buffer bypass implemented with multiplexers, forwarding the weights from one slab directly to the next while bypassing redundant slab-local weight buffers. When fused, adjacent slabs execute a single tile as a larger logical systolic array. Multiple fused slab groups may operate concurrently, each executing a different tile of the same \ac{GEMM}.

\subsubsection{Buffering Organization.}

SISA employs a hierarchical buffering organization composed of a global buffer, slab-local buffers, and a dedicated global output buffer. 
All buffers employ double buffering to overlap data movement with computation.
The global buffer stores activations and weights, while the output buffer collects the results produced by the slabs before they are written back to memory.

\textit{Global buffer organization.}
The global buffer is implemented as a collection of partitioned SRAM banks. Each bank is dedicated to providing activations or weights to a specific set of local slab buffers. Thanks to the regular data access patterns of \ac{SA}-based \ac{GEMM} execution, this organization enables building larger SRAM capacity while maintaining bounded access latency. Each bank has a port width of multiple elements, allowing it to write to several local buffers per cycle and increasing effective bandwidth with limited latency overhead.

\textit{Output buffer organization.} 
A multi-banked output buffer collects the results produced by the slabs and supports both independent and fused slab execution.

\textit{Local slab buffers.}
Each slab contains small local buffers for activations and weights placed next to the array boundary to directly feed the \acp{PE}. 
Each \ac{PE} along the upper boundary of the slab receives data from a dedicated pair of banks.

\subsubsection{Power-Gating.}
Power management in SISA is performed at the granularity of individual slabs. Slab-local buffers can be selectively disabled, and entire slabs can be power-gated when not required for execution. This coarse-grain power-gating strategy minimizes idle energy consumption.

\begin{figure*}[t]
    \centering
    \begin{subfigure}{0.48\textwidth}
        \centering
        \includegraphics[width=\linewidth]{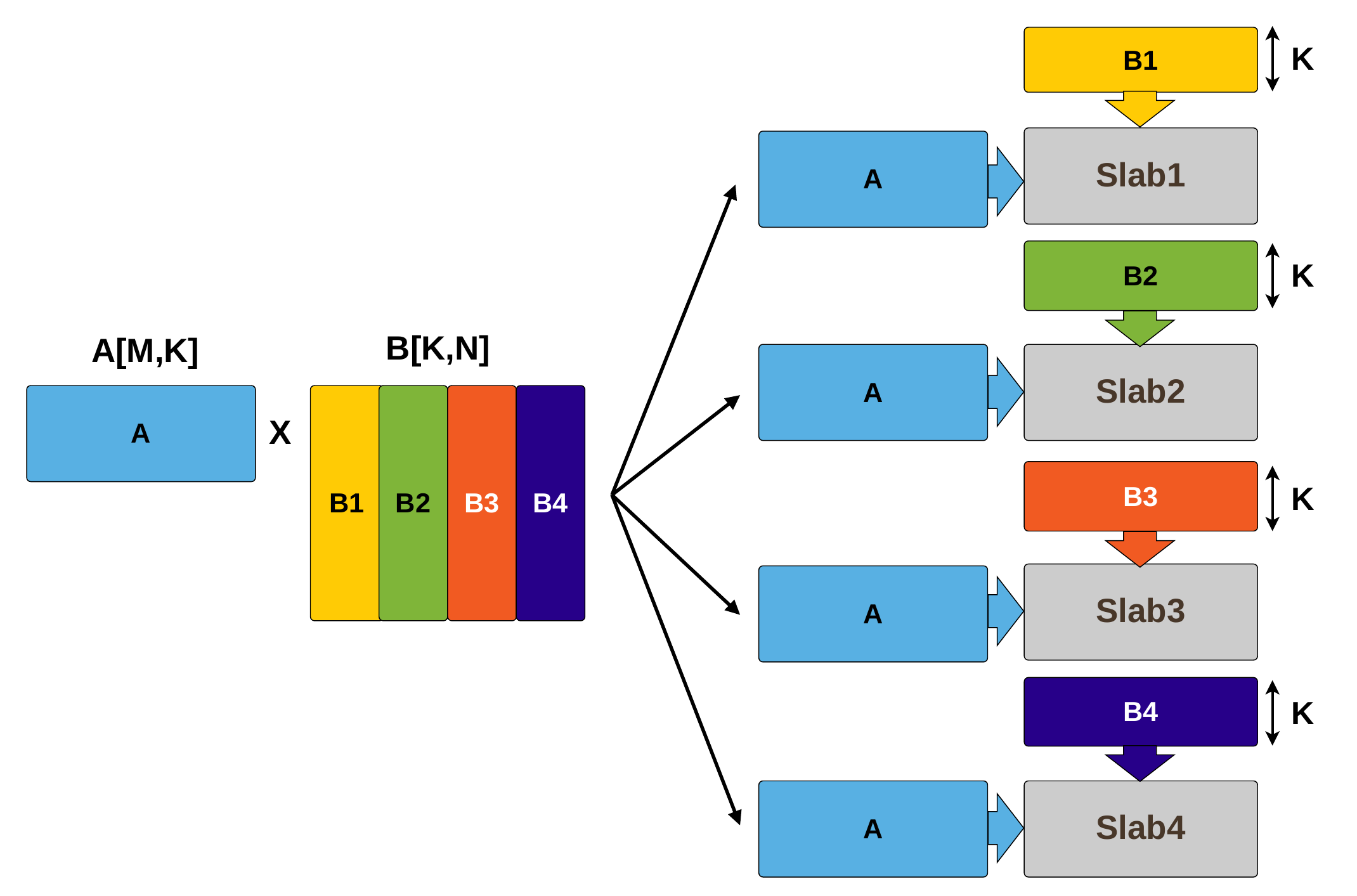}
        \caption{}
        \label{fig:tiling}
    \end{subfigure}\hfill
    \begin{subfigure}{0.48\textwidth}
        \centering
        \includegraphics[width=\linewidth]{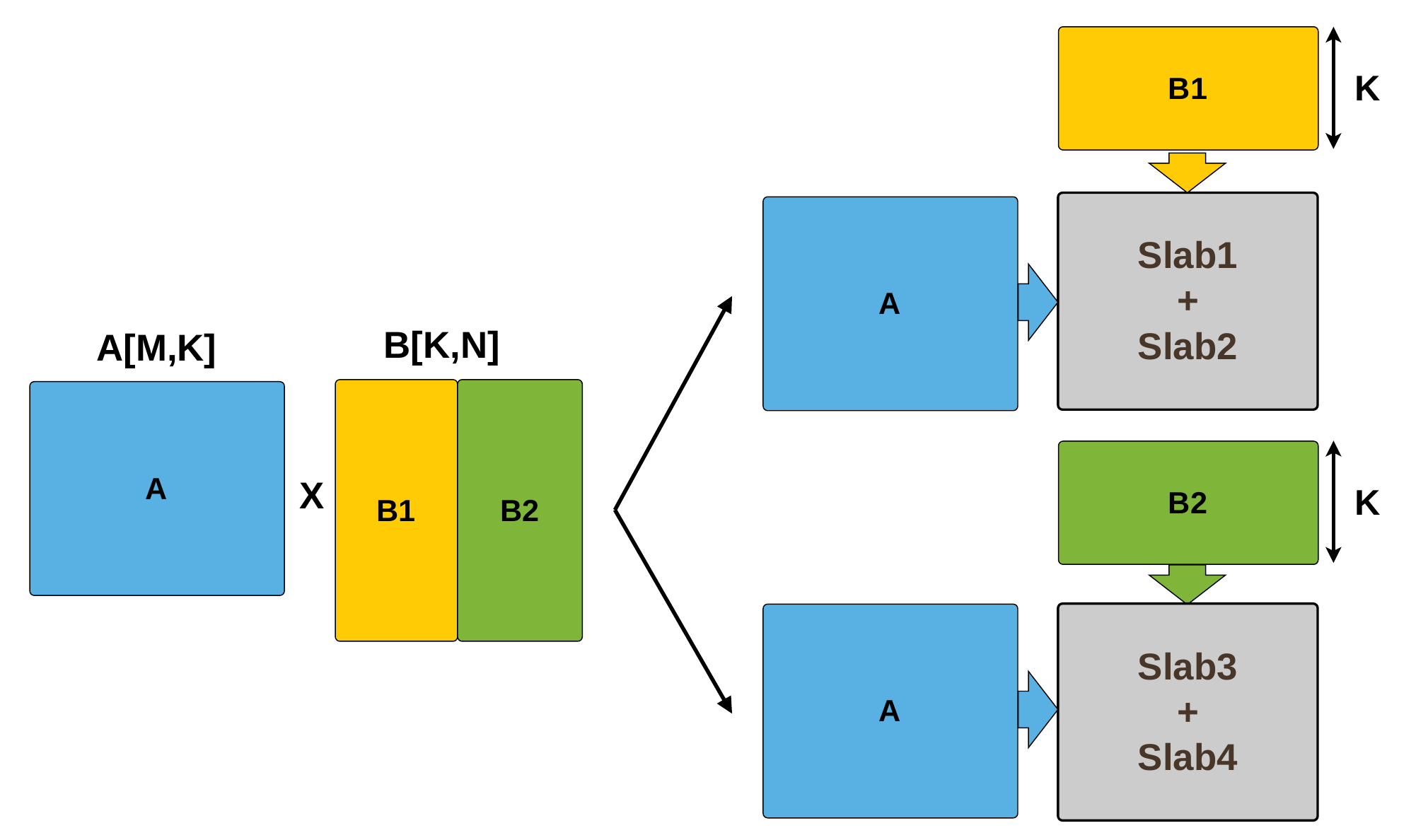}
        \caption{}
        \label{fig:tiling_fused}
    \end{subfigure}

    \vspace{0.5em}

    \begin{subfigure}{0.5\textwidth}
         \centering
        \includegraphics[width=\linewidth]{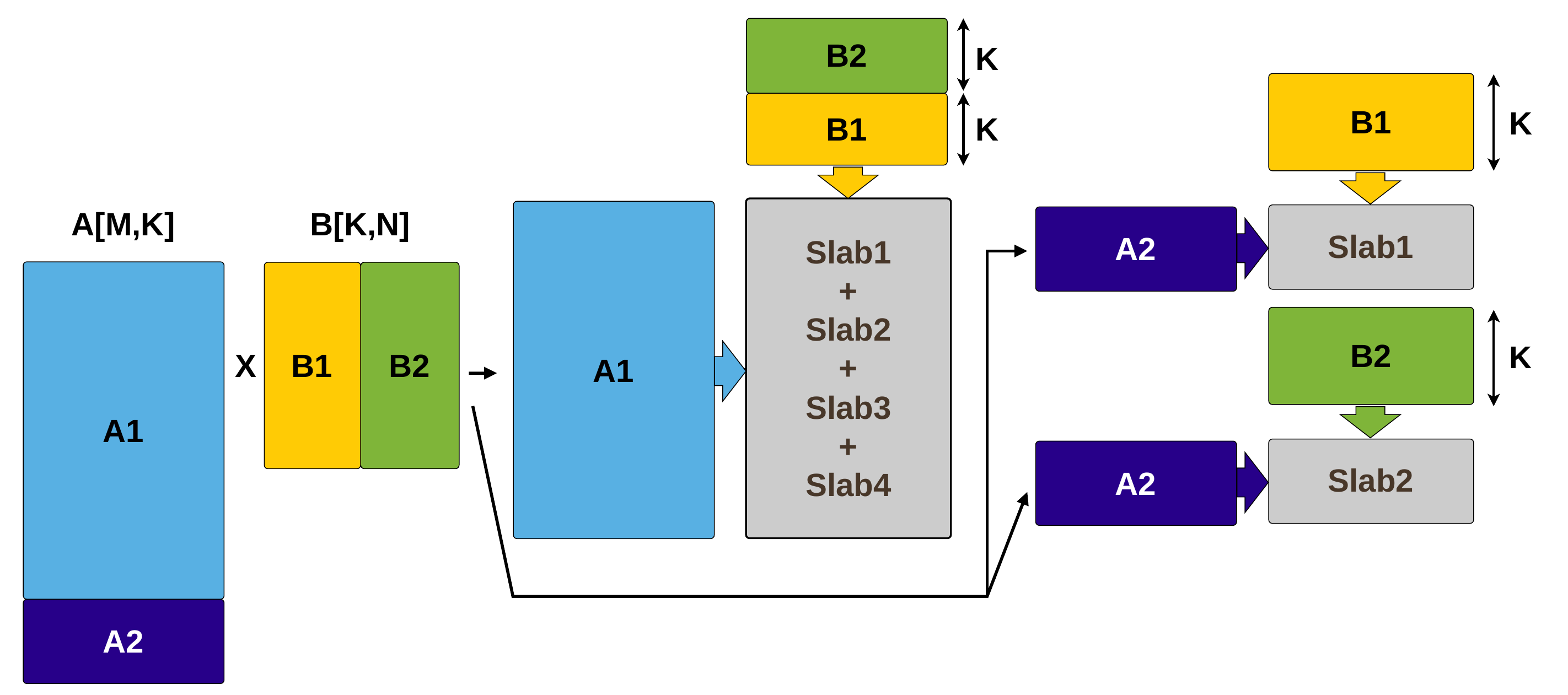}
        \caption{}
        \label{fig:tiling_monolithic}
        
    \end{subfigure}\hfill
    \begin{subfigure}{0.48\textwidth}
       \centering
        \includegraphics[width=\linewidth]{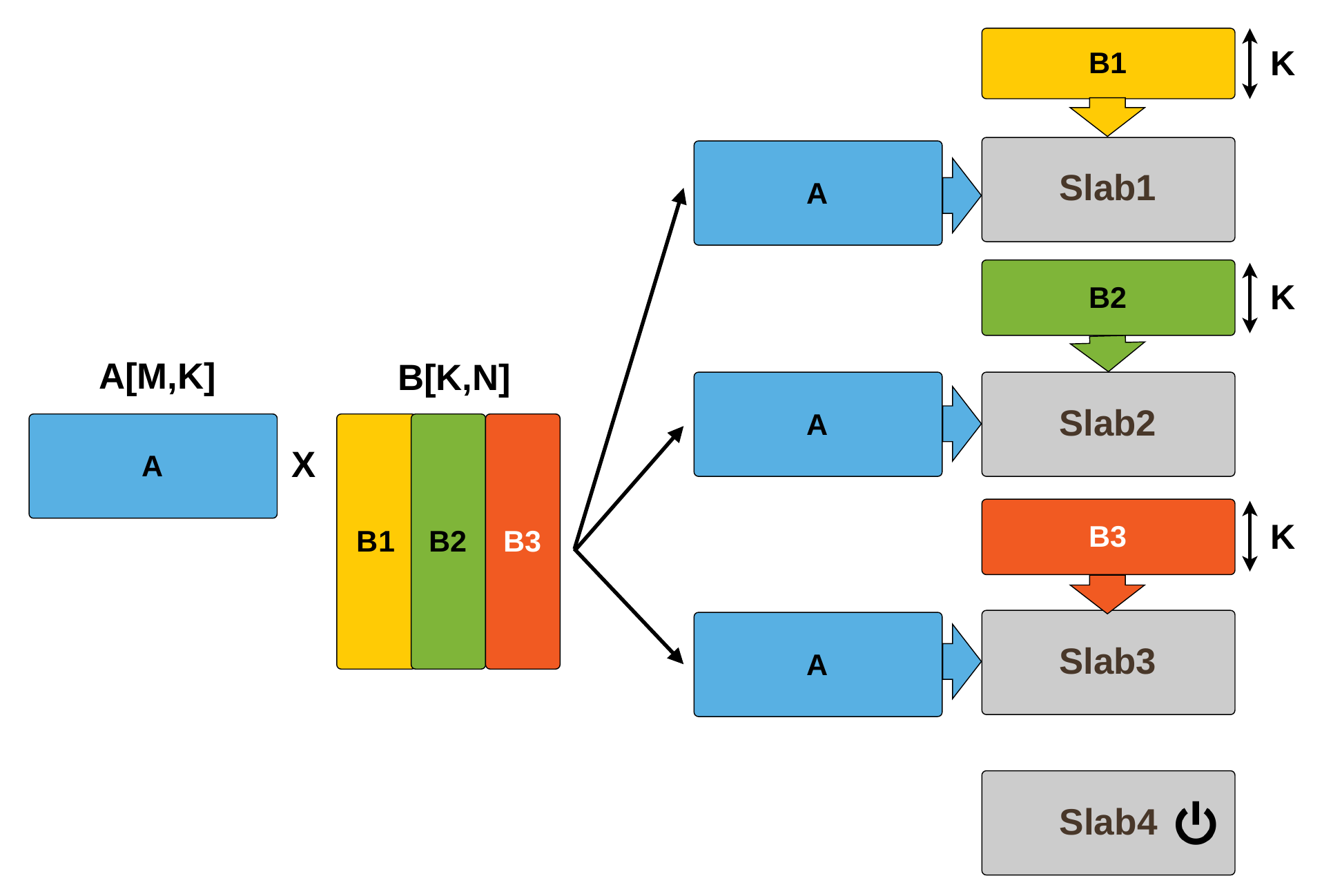}
        \caption{}
        \label{fig:tiling_power_gate}
    \end{subfigure}

    \caption{Tiling and execution strategies in SISA for different GEMM shapes. (a) Independent slab execution for small-$M$ GEMMs, distributing tiles across slabs along $N$. (b) Slab fusion for intermediate-$M$ GEMMs, forming larger logical \acp{SA} matching tile height. (c) Monolithic execution for large-$M$ GEMMs using the full array. (d) Slab-level power-gating disables slabs when parallelism is limited.}

    \label{fig:tiling_strategies}
    \vspace{-12pt}
\end{figure*}

\subsection{Tiling and Scheduling}
\label{subsec:sisa-tiling}

SISA decomposes each GEMM into tiles that map naturally onto slab configurations. 
Given $C[M,N] = A[M,K] \times B[K,N]$, SISA tiles the $M$, $N$, and $K$ dimensions to satisfy on-chip storage constraints while maximizing data reuse and \ac{SA} utilization. 
Tiles are statically scheduled across slabs (or fused slab groups), enabling concurrent execution when resources are available. When tiling along $K$ is required, partial tiles are processed sequentially and accumulated before advancing to the next output tile. 
Figure~\ref{fig:tiling_strategies} illustrates the execution strategies across different GEMM shapes.

\textbf{M $\leq$ slab height.}
When $M$ is smaller than the slab height, SISA distributes computation along the $N$ dimension, as shown in Figure~\ref{fig:tiling}. 
The input matrix $A$ is loaded once into on-chip memory and reused across tiles, while the weight matrix $B$ is partitioned along $N$ into tiles matching the slab width. 
Each $B$ tile is assigned to a slab, allowing multiple tiles to execute in parallel. If the number of tiles is smaller than the available slabs, unused slabs are power-gated, as illustrated in Figure~\ref{fig:tiling_power_gate}. If tiles exceed on-chip storage capacity, SISA further partitions the computation along $K$ while keeping $A$ resident to maximize reuse.

\textbf{Slab height $<$ M $\leq$ array height.}
When $M$ exceeds the slab height but remains smaller than the full systolic array height, SISA fuses multiple slabs along $M$ to form a taller logical slab, illustrated in Figure~\ref{fig:tiling_fused}. 
The fused slabs execute tiles whose $M$ dimension matches the fused height. 
After determining the fusion configuration, SISA applies the same tiling strategy along $N$. Each fused slab group processes one tile, enabling parallel execution across groups. Because fusion reduces the number of independent slab groups, fewer tiles may execute concurrently and remaining tiles are processed in subsequent iterations.

\textbf{M $>$ array height.}
When $M$ exceeds the full array height, SISA executes the GEMM using a fully fused configuration equivalent to a monolithic systolic array, shown in Figure~\ref{fig:tiling_monolithic}. 
The computation is decomposed into a main tile spanning the full array height. After completing this tile, residual tiles are executed using the appropriate slab configuration, either by fusing slabs to match the residual $M$ dimension or by distributing the computation across independent slabs.

%% file: sections/04_results.tex
\section{Evaluation}
\label{sec:evaluation}

\subsection{Experimental Setup}
\label{sec:setup}

For area and energy evaluation, we model SISA using RTL synthesis and SRAM estimation tools. The design is synthesized with Cadence Genus 18.14 targeting a 28\,nm technology node with the ASAP7 cell library~\cite{asap7_cell_library} at 1\,GHz. On-chip buffers are modeled using CACTI~\cite{cacti7,cacti_og}. For performance evaluation, we extend SCALE-Sim~\cite{scalesim} to support the architectural features required by our design.

We compare SISA against two representative \ac{SA} baselines: Google's TPUv4~\cite{tpuv4}, representing a monolithic \ac{SA}, and ReDas~\cite{ReDAs}, a reconfigurable \ac{SA} that  reports improvements over prior SOTA designs discussed in Section~\ref{sec:related}. All baselines are modeled according to their published configurations using the same RTL+CACTI methodology to ensure fairness.
CACTI-based SRAM estimations were validated against the detailed memory area numbers reported in ReDas, obtained using TSMC's 28\,nm memory compiler. The relative error is only 0.27\%.

SISA is evaluated using a $128 \times 128$ \ac{SA}, consistent with the configurations adopted in TPUv4 and ReDas. We do not model the roundabout interconnect overhead in ReDas, which benefits this baseline. Further architectural details of the evaluated designs are provided in Section~\ref{sec:sisa_design}.

\begin{table}[t]
    \centering
    
    \caption{Dimensions of unique \ac{GEMM} operations across models, reported as triples $(M,N,K)$. $m$ denotes the sequence length in the prefill phase or batch size in the decode phase. }
    \label{tab:llm-gemms}
    \begin{tabular}{c c c c c}
    \toprule
    ID & Qwen2.5-0.5B & Qwen2.5-1.5B & Llama3.2-3B & Qwen2.5-7B \\
    \midrule
    0 & (m,896,896)    & (m,1536,1536) & (m,3072,3072) & (m,3584,3584) \\
    1 & (m,128,896)    & (m,356,1536) & (m,1024,3072) & (m,512,3584) \\
    2 & (m,4864,896)   & (m,8960,1536) & (m,8192,3072) & (m,3584,18944) \\
    3 & (m,896,4864)   & (m,1536,8960) & (m,3072,8192) & (m,18944,3584) \\
    4 & (m,151936,896) & (m,151936,1536) & (m,128256,3072) &(m,152064,3584) \\
    \bottomrule
    \end{tabular}
\end{table}

As \ac{GEMM} is the dominant operation in \ac{LLM} execution and since SISA is an architecture to optimize \ac{GEMM}, we perform this evaluation on the \ac{GEMM} operations present in modern relevant \acp{LLM} of different size: Qwen2.5\footnote{https://huggingface.co/collections/Qwen/qwen25} 0.5B, 1.5B, and 7B, as well as Llama3.2 3B\footnote{https://huggingface.co/meta-llama/Llama-3.2-3B-Instruct}. We extracted the model information from the pre-trained models available in HuggingFace. The \ac{LLM} models are composed of many layers but they are just replications of common blocks. In Table~\ref{tab:llm-gemms} we report the unique \ac{GEMM} sizes as triples $(M,N,K)$.

\subsection{Designing a $128 \times 128$ PE 8-Slab SISA}
\label{sec:sisa_design}
\input{tables/sisa_config}

\subsubsection{Slabs.}
The slab height of 16 is primarily chosen to match practical off-chip bandwidth limits. Concurrently streaming activations and weights for eight slabs at BF16 precision requires approximately 2.3\,TB/s, which aligns with the capabilities of modern HBM4 systems (up to $\sim$2.8\,TB/s~\cite{micron_hbm4}). Finer-grain partitioning would exceed feasible bandwidth constraints and is therefore not considered.

\subsubsection{Memory Organization.}
For a fair comparison, we allocate the same total on-chip memory capacity across TPU, ReDas, and SISA, organized according to each architecture. 
The TPU-like baseline employs two 4\,MB buffers (activation and weight) and one 2\,MB output buffer. 
ReDas distributes four 2.5\,MB buffers around the \ac{SA}, following its multi-mode organization. 
SISA preserves the same total memory budget using 8\,MB global activation and weight buffer and 2\,MB of output buffer.

\subsubsection{Area and Energy Breakdown.}

Table~\ref{tab:sisa_config} reports the post-synthesis area and per-cycle static energy of SISA at 1\,GHz. The $128\times128$ BF16 \ac{SA} dominates both area (192.91mm$^2$) and energy (21.60nJ/cycle), while the memory hierarchy contributes 6.59nJ/cycle from leakage. Dynamic SRAM and DRAM energies are modeled separately using per-access energy parameters and are accounted for during workload execution.

\subsection{SISA Evaluation}
\label{sec:sisa_eval}

\begin{figure*}[!t]
  \centering
  \includegraphics[width=\textwidth,height=0.40\textheight,keepaspectratio]{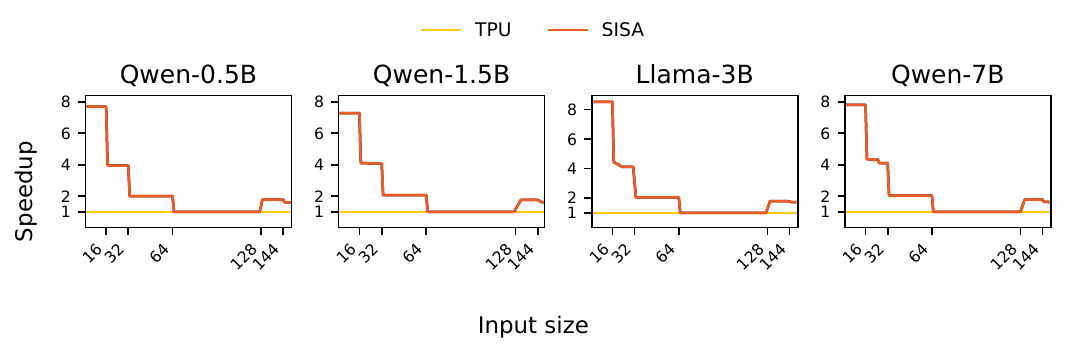}
  \caption{Speedup of SISA compared to TPU. Higher is better.}
  \label{fig:sisa_vs_tpu_speedup}
  \vspace{-12pt}
\end{figure*}

\begin{figure*}[!t]
  \centering
  \includegraphics[width=\textwidth,height=0.40\textheight,keepaspectratio]{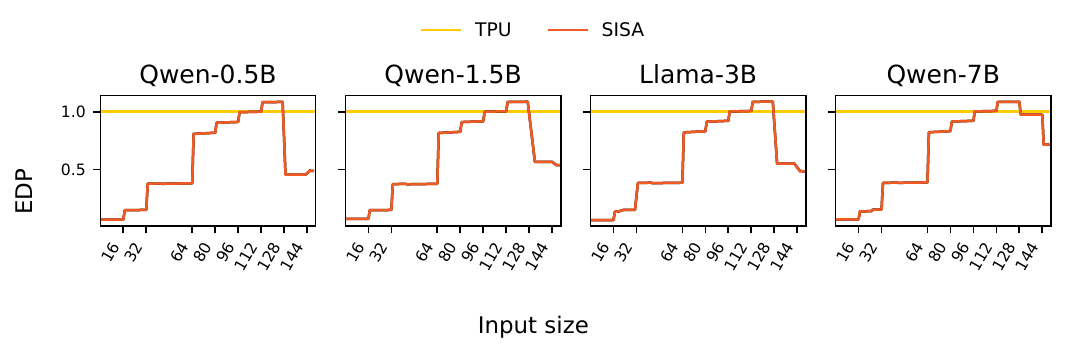}
  \caption{Normalized EDP of SISA compared to TPU. Lower is better.}
  \label{fig:sisa_vs_tpu_edp}
  \vspace{-12pt}
\end{figure*}

Figures~\ref{fig:sisa_vs_tpu_speedup} and~\ref{fig:sisa_vs_tpu_edp} report the speedup and normalized EDP of SISA over a monolithic TPU-like systolic array for sequence lengths $m$ from 1 to 150, covering the different operating modes of SISA and their trade-offs. Each point aggregates the execution of the linear layers listed in Table~\ref{tab:llm-gemms}, scaled by the number of times each layer appears in the model. In the prefill phase, this corresponds to all \acp{GEMM} required to process a prompt of length $m$, while in the decode phase it corresponds to all \acp{GEMM} required to generate one token with batch size $m$.

\textbf{Small sequence lengths ($m \leq 16$).}

For small $m$ dimensions, SISA executes slabs independently, exposing fine-grain parallelism and improving PE utilization. In this regime, SISA achieves up to 8.52$\times$ speedup and reduces EDP by up to 93\% compared to TPU. This improvement comes from eliminating the severe under-utilization of monolithic arrays when mapping skewed GEMMs. The speedup slightly exceeds the 8$\times$ parallelism provided by the eight slabs because a monolithic $128$-PE array must drain outputs across the full array height even when the output dimension is small. In contrast, slab-level execution avoids this overhead and writes results directly to the global buffer.

\textbf{Intermediate sequence lengths ($16 < m \leq 64$).}
As $m$ increases, SISA progressively fuses slabs into larger logical arrays ($32\times128$ and $64\times128$), trading parallelism for improved data reuse. In these configurations, SISA achieves up to 4.12$\times$ and 2.06$\times$ speedup, with EDP reductions of up to 85\% and 65\%, respectively, as performance gradually approaches that of a monolithic array.

\textbf{Large sequence lengths ($m > 64$).}
For $64 < m \leq 128$, both architectures operate as a fully fused array and achieve comparable performance. In this range, SISA enables slab-level power gating when the effective $M$ dimension is underutilized, reducing EDP by up to 18\%. When the array is fully utilized ($112 < m \leq 128$), SISA incurs an 8.47\% EDP overhead due to its distributed memory hierarchy and additional local buffers and multiplexers. For $m > 128$, after executing the full $128\times128$ tile, SISA processes the remaining partial tiles using independent slabs, yielding up to 1.79$\times$ additional speedup and 46\% EDP reduction.

\textbf{Area Comparison.}
Table~\ref{tab:sisa_config} shows that the systolic array dominates total chip area. Relative to a BF16 TPU with the same $128\times128$ array, SISA introduces a 3\% PE-array overhead for slab-level power gating~\cite{xue2025regate-powergating-overhead}, corresponding to 2.7\% of total chip area. Using different bank sizes, wider port widths, and slab-local buffers slightly increases SRAM area compared to the TPU design, contributing an additional 2.74\% of total area. Overall, SISA increases total chip area by about 5.44\%.

In summary, SISA delivers substantial gains in the under-utilized regime while maintaining comparable performance to monolithic arrays at large sequence lengths, with only modest area overhead.

\subsection{SISA versus Alternative Reconfigurable SA}

\begin{figure*}[!t]
  \centering
  \includegraphics[width=\textwidth,height=0.40\textheight,keepaspectratio]{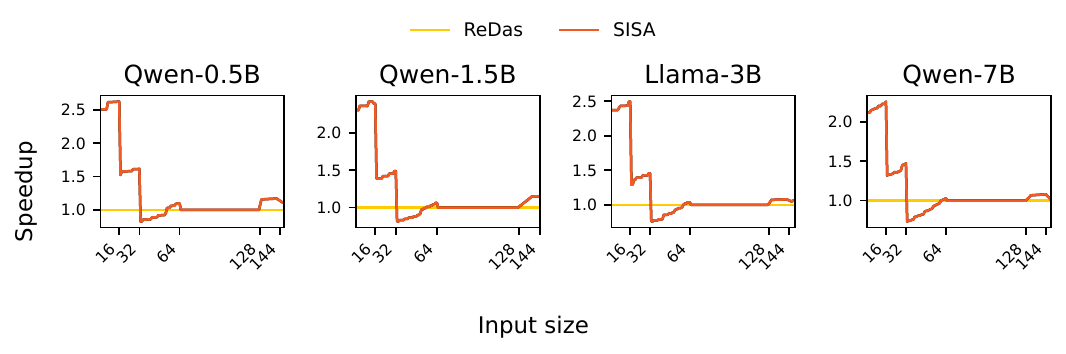}
  \caption{Speedup of SISA compared to ReDas.}
  \label{fig:sisa_vs_redas_speedup}
  \vspace{-12pt}
\end{figure*}
\begin{figure*}[t]
  \centering
  \includegraphics[width=\textwidth,height=0.40\textheight,keepaspectratio]{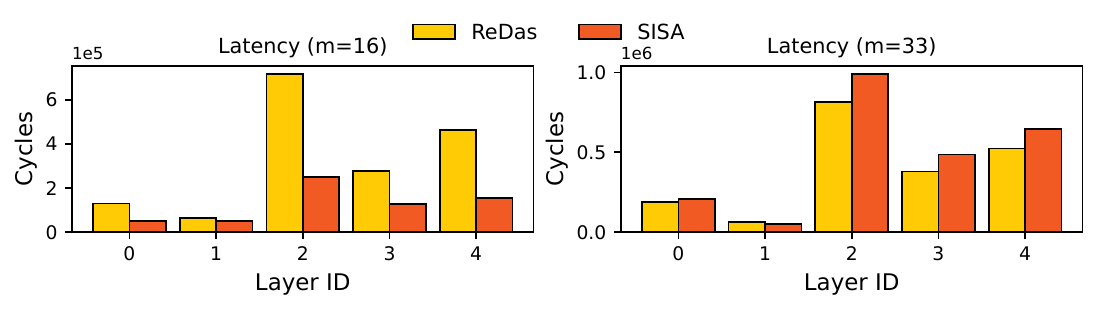}
  \caption{Latency of Qwen2.5-0.5B linear layers at $m=16$ (best case) and $m=33$ (worst case) for SISA, weighted by layer occurrence. Lower is better.}
  \label{fig:layer_latency_comp}
  \vspace{-12pt}
\end{figure*}

Figure~\ref{fig:sisa_vs_redas_speedup} shows the speedup of SISA over ReDas across the same sequence-length sweep. We compare performance against ReDas using an extended version of SCALE-Sim, as described in Section~\ref{sec:setup}. To avoid penalizing ReDas, the model abstracts certain control and data-movement overheads, making the comparison favorable to ReDas in terms of latency. For this reason, we report performance only and omit EDP results. The relative area and energy overheads of the two designs are analyzed later in this section.

\textbf{Small sequence lengths ($1 \leq m \leq 32$).}
In this regime, SISA outperforms ReDas. Operating in independent slab mode, SISA maximizes PE utilization and achieves up to 2.61$\times$ and 1.61$\times$ speedup in the $16\times128$ and $32\times128$ configurations, respectively. Although ReDas reshapes its array, its coarse granularity limits effective PE activation for skewed GEMMs, resulting in lower utilization.

\textbf{Large sequence lengths ($m \geq 33$).}
As $m$ increases, ReDas performs best in the mid-range ($m\approx33$–50) across all models, where its reshaping aligns well with the workload. Beyond $m\approx50$, SISA regains advantage for Qwen2.5-0.5B and Qwen2.5-1.5B, whose layer dimensions better match slab sizes, reducing residual-tile overhead that continues to affect ReDas. For larger models, the higher compute volume amortizes residual tiles, allowing ReDas to maintain a modest advantage, although SISA underperforms by at most 1.36$\times$. At $m=64$, SISA slightly recovers despite using the same PE budget: ReDas forms a $64\times256$ array, while SISA operates as two $64\times128$ slabs, reducing fill-and-drain overhead. For $64 \leq m \leq 128$, both architectures operate in effectively monolithic configurations and achieve comparable performance. Beyond $m>128$, SISA again benefits from slab-level execution when processing residual tiles, achieving up to 1.17$\times$ speedup over ReDas.

\textbf{Case Study: Qwen2.5-0.5B ($m=16$ vs.\ $m=33$).}
As shown in Figure~\ref{fig:sisa_vs_redas_speedup}, SISA outperforms ReDas across most input sizes in the $1$–$150$ range, while the performance loss in the few unfavorable cases remains limited. To illustrate these trends, we analyze two representative points: the best-case configuration ($m=16$) and the worst-case configuration ($m=33$). Table~\ref{tab:llm-gemms} lists the linear layers of Qwen2.5-0.5B, and Figure~\ref{fig:layer_latency_comp} shows their latency in cycles weighted by layer occurrence for these two cases. The overall results are presented earlier in this section.

For $m=16$, SISA operates in an $8\times(16\times128)$ configuration, while ReDas forms a $16\times448$ array. In this setting, SISA achieves higher effective utilization and reduced tail inefficiencies. Additionally, 44\% of execution occurs with one or more slabs power-gated, improving energy efficiency without affecting latency. Although smaller than the final projection layer, Layer~2 ($M=16$, $N=4864$, $K=896$) is repeated 48 times per block and therefore dominates the total latency.

At $m=33$, SISA transitions to a fused $2\times64\times128$ configuration, while ReDas reshapes into a $32\times384$ array, activating more PEs and achieving its peak advantage. For Layer~1 ($M=33$, $N=896$, $K=896$), SISA still attains slightly higher utilization due to better slab alignment, partially offsetting the additional PE availability of ReDas.

\textbf{Area and Energy Comparison.}
ReDas exhibits a slightly lower memory overhead, with SRAM accounting for about 0.4\% less of the total chip area compared to SISA. However, when considering the \ac{PE}-array, ReDas reports a 70\% per-PE area overhead and a $2.49\times$ increase power for an INT8 design. When scaling the design from INT8 to BF16, the PE area increases due to the wider datapath. In our design, the PE array dominates chip area (87.2\%), while on-chip SRAM accounts for 12.8\%. Therefore, increasing PE complexity has a much larger impact on total area and energy. Although the relative overhead of ReDas may decrease in BF16, it still introduces additional logic inside each PE. In contrast, SISA preserves a simple PE design and concentrates overhead in the memory hierarchy, resulting in only a 5.44\% total chip area increase and a lower overall overhead than ReDas.

Overall, SISA achieves higher performance in most regimes and comparable performance otherwise, while incurring lower area and energy overhead.

%% file: tables/sisa_config.tex
\begin{table}[t]
\caption{SISA Configuration, Area, and Per-Cycle Static Energy (1GHz)}
\label{tab:sisa_config}
\centering
\begin{tabular}{lccc}
\toprule
Parameter & Value & Area & Energy / Cycle (nJ) \\ 
\midrule
\ac{SA} Size              & $128 \times 128$ & 192.91 mm$^2$ & 21.60 \\
Global Buffer Size        & 8\,MB            & 22.45 mm$^2$  & 5.22 \\
Slabs Buffer Sizes         & 8\,KB, 64\,KB    & 0.30 mm$^2$   & 0.12 \\
Output Buffer Size        & 2\,MB            & 5.61 mm$^2$   & 1.25 \\
\midrule
\textbf{Total}            & --               & \textbf{221.27 mm$^2$} & \textbf{28.19nJ} \\
\bottomrule
\end{tabular}
\end{table}

%% file: sections/05_conclusion.tex
\section{Conclusion}

In this work we proposed SISA, a new SA architecture based on slabs of PEs which is able to achieve higher speedup and better EDP than SOTA flexible and monolithic SA architectures for operations with highly skewed matrices. This design is particularly effective for small- to medium-batch decoding, the most common setting in interactive applications, and for latency-sensitive prefill like in chatbot workloads with short prompts (median $\approx$12 tokens).
With a minimal area overhead over the baseline TPUv4, SISA delivers up to $8.52\times$ speedup and 93\% EDP reduction compared to that baseline, while in the worst case achieving comparable performance with only 8.47\% higher EDP. When compared to the best alternative SOTA, SISA performs better in most cases, achieving up to $2.61\times$ speedup while also exhibiting lower area overhead.

%% file: sections/06_acknowledgments.tex
\begin{credits}
\subsubsection{\ackname} This work was supported in part by SSF, Swedish Foundation for Strategic Research, under grant number SIP21-0087.
%
\end{credits}

%% file: bibliography.bib
@ARTICLE{ReDAs,
    author={Han, Meng and Wang, Liang and Xiao, Limin and Cai, Tianhao and Wang, Zeyu and Xu, Xiangrong and Zhang, Chenhao},
    journal={ IEEE Transactions on Computers },
    title={{ ReDas: A Lightweight Architecture for Supporting Fine-Grained Reshaping and Multiple Dataflows on Systolic Array }},
    year={2024},
    volume={73},
    number={08},
    ISSN={1557-9956},
    pages={1997-2011},
    doi={10.1109/TC.2024.3398500},
    url = {https://doi.ieeecomputersociety.org/10.1109/TC.2024.3398500},
    publisher={IEEE Computer Society},
    address={Los Alamitos, CA, USA},
    month=aug
}

@misc{FlexSA,
      title={FlexSA: Flexible Systolic Array Architecture for Efficient Pruned DNN Model Training}, 
      author={Sangkug Lym and Mattan Erez},
      year={2020},
      eprint={2004.13027},
      archivePrefix={arXiv},
      primaryClass={cs.LG},
      url={https://arxiv.org/abs/2004.13027}, 
}

@inproceedings{SARA,
    author = {Samajdar, Ananda and Qin, Eric and Pellauer, Michael and Krishna, Tushar},
    title = {Self adaptive reconfigurable arrays (SARA): learning flexible GEMM accelerator configuration and mapping-space using ML},
    year = {2022},
    isbn = {9781450391429},
    publisher = {Association for Computing Machinery},
    address = {New York, NY, USA},
    url = {https://doi.org/10.1145/3489517.3530506},
    doi = {10.1145/3489517.3530506},
    booktitle = {Proceedings of the 59th ACM/IEEE Design Automation Conference},
    pages = {583–588},
    numpages = {6},
    location = {San Francisco, California},
    series = {DAC '22}
}

@ARTICLE{DyNNamic,
  author={Hanson, Edward and Li, Shiyu and Qian, Xuehai and Li, Hai Helen and Chen, Yiran},
  journal={IEEE Transactions on Computers}, 
  title={DyNNamic: Dynamically Reshaping, High Data-Reuse Accelerator for Compact DNNs}, 
  year={2023},
  volume={72},
  number={3},
  pages={880-892},
  doi={10.1109/TC.2022.3184272}}

@INPROCEEDINGS{Planaria,
  author={Ghodrati, Soroush and Ahn, Byung Hoon and Kyung Kim, Joon and Kinzer, Sean and Yatham, Brahmendra Reddy and Alla, Navateja and Sharma, Hardik and Alian, Mohammad and Ebrahimi, Eiman and Kim, Nam Sung and Young, Cliff and Esmaeilzadeh, Hadi},
  booktitle={2020 53rd Annual IEEE/ACM International Symposium on Microarchitecture (MICRO)}, 
  title={Planaria: Dynamic Architecture Fission for Spatial Multi-Tenant Acceleration of Deep Neural Networks}, 
  year={2020},
  volume={},
  number={},
  pages={681-697},
  doi={10.1109/MICRO50266.2020.00062}}

@ARTICLE{savector,
  author={Choi, Sangun and Park, Seongjun and Park, Jaeyong and Kim, Jongmin and Koo, Gunjae and Hong, Seokin and Yoon, Myung Kuk and Oh, Yunho},
  journal={IEEE Access}, 
  title={SAVector: Vectored Systolic Arrays}, 
  year={2024},
  volume={12},
  number={},
  pages={44446-44461},
  doi={10.1109/ACCESS.2024.3380433}}

@article{sosa,
author = {Y\"{u}z\"{u}g\"{u}ler, Ahmet Caner and S\"{o}nmez, Canberk and Drumond, Mario and Oh, Yunho and Falsafi, Babak and Frossard, Pascal},
title = {Scale-out Systolic Arrays},
year = {2023},
issue_date = {June 2023},
publisher = {Association for Computing Machinery},
address = {New York, NY, USA},
volume = {20},
number = {2},
issn = {1544-3566},
url = {https://doi.org/10.1145/3572917},
doi = {10.1145/3572917},
journal = {ACM Trans. Archit. Code Optim.},
month = mar,
articleno = {27},
numpages = {25}
}

@INPROCEEDINGS{tpuv4,
  author={Jouppi, Norman P. and Hyun Yoon, Doe and Ashcraft, Matthew and Gottscho, Mark and Jablin, Thomas B. and Kurian, George and Laudon, James and Li, Sheng and Ma, Peter and Ma, Xiaoyu and Norrie, Thomas and Patil, Nishant and Prasad, Sushma and Young, Cliff and Zhou, Zongwei and Patterson, David},
  booktitle={2021 ACM/IEEE 48th Annual International Symposium on Computer Architecture (ISCA)}, 
  title={Ten Lessons From Three Generations Shaped Google’s TPUv4i : Industrial Product}, 
  year={2021},
  volume={},
  number={},
  pages={1-14},
  doi={10.1109/ISCA52012.2021.00010}
}

@inbook{gemmini,
author = {Genc, Hasan and Kim, Seah and Amid, Alon and Haj-Ali, Ameer and Iyer, Vighnesh and Prakash, Pranav and Zhao, Jerry and Grubb, Daniel and Liew, Harrison and Mao, Howard and Ou, Albert and Schmidt, Colin and Steffl, Samuel and Wright, John and Stoica, Ion and Ragan-Kelley, Jonathan and Asanovic, Krste and Nikolic, Borivoje and Shao, Yakun Sophia},
title = {Gemmini: Enabling Systematic Deep-Learning Architecture Evaluation via Full-Stack Integration},
year = {2022},
isbn = {9781665432740},
publisher = {IEEE Press},
url = {https://doi.org/10.1109/DAC18074.2021.9586216},
booktitle = {Proceedings of the 58th Annual ACM/IEEE Design Automation Conference},
pages = {769–774},
numpages = {6}
}

@INPROCEEDINGS{scalesim,
  author={Raj, Ritik and Banerjee, Sarbartha and Chandra, Nikhil and Wan, Zishen and Tong, Jianming and Samajdhar, Ananda and Krishna, Tushar},
  booktitle={2025 IEEE International Symposium on Performance Analysis of Systems and Software (ISPASS)}, 
  title={SCALE-Sim V3: a Modular Cycle-Accurate Systolic Accelerator Simulator for End-To-End System Analysis}, 
  year={2025},
  volume={},
  number={},
  pages={186-200},
  doi={10.1109/ISPASS64960.2025.00026}}

@inproceedings{asap7_cell_library,
  title={ASAP7 predictive design kit development and cell design technology co-optimization},
  author={Vashishtha, Vinay and Vangala, Manoj and Clark, Lawrence T},
  booktitle={2017 IEEE/ACM International Conference on Computer-Aided Design (ICCAD)},
  pages={992--998},
  year={2017},
  organization={IEEE}
}

@article{cacti_og,
  title={CACTI: An enhanced cache access and cycle time model},
  author={Wilton, Steven JE and Jouppi, Norman P},
  journal={IEEE Journal of solid-state circuits},
  volume={31},
  number={5},
  pages={677--688},
  year={1996},
  publisher={IEEE}
}

@article{cacti7,
  title={CACTI 7: New tools for interconnect exploration in innovative off-chip memories},
  author={Balasubramonian, Rajeev and Kahng, Andrew B and Muralimanohar, Naveen and Shafiee, Ali and Srinivas, Vaishnav},
  journal={ACM Transactions on Architecture and Code Optimization (TACO)},
  volume={14},
  number={2},
  pages={1--25},
  year={2017},
  publisher={ACM New York, NY, USA}
}

@book{kung_why_sa,
  title={Why systolic architecture?},
  author={Kung, Hsiang-Tsung},
  year={1982},
  publisher={Design Research Center, Carnegie-Mellon University Pittsburgh, PA, USA}
}

@inproceedings{mutlu_tpu_utilization,
  title={Google neural network models for edge devices: Analyzing and mitigating machine learning inference bottlenecks},
  author={Boroumand, Amirali and Ghose, Saugata and Akin, Berkin and Narayanaswami, Ravi and Oliveira, Geraldo F and Ma, Xiaoyu and Shiu, Eric and Mutlu, Onur},
  booktitle={2021 30th International Conference on Parallel Architectures and Compilation Techniques (PACT)},
  pages={159--172},
  year={2021},
  organization={IEEE}
}

@misc{micron_hbm4,
  title = {Building AI, region by region: Why memory and storage define the next decade},
  author = {Viral Gosalia},
  year = {2025},
  publisher = {Micron},
  howpublished = {\url{https://www.micron.com/about/blog/applications/ai/building-ai-region-by-region-why-memory-and-storage-define-the-next-decade}},
}

@inproceedings{halo,
  title={HALO: Hybrid Systolic Arrays via Logical Partitioning for Acceleration of Complex-Valued Neural Networks},
  author={Yi, Ji Yeong and Jeong, Eunbi and Yum, SungHee and Rhee, Jane and Choi, Sangun and Koo, Gunjae and Oh, Yunho and Yoon, Myung Kuk},
  booktitle={2025 IEEE International Symposium on Workload Characterization (IISWC)},
  pages={206--218},
  year={2025},
  organization={IEEE}
}

@misc{wang2025llmservingoptimizationvariable,
      title={LLM Serving Optimization with Variable Prefill and Decode Lengths}, 
      author={Meixuan Wang and Yinyu Ye and Zijie Zhou},
      year={2025},
      eprint={2508.06133},
      archivePrefix={arXiv},
      primaryClass={math.OC},
      url={https://arxiv.org/abs/2508.06133}, 
}

@misc{ sarathi_serve,
      title={Taming Throughput-Latency Tradeoff in LLM Inference with Sarathi-Serve}, 
      author={Amey Agrawal and Nitin Kedia and Ashish Panwar and Jayashree Mohan and Nipun Kwatra and Bhargav S. Gulavani and Alexey Tumanov and Ramachandran Ramjee},
      year={2024},
      eprint={2403.02310},
      archivePrefix={arXiv},
      primaryClass={cs.LG},
      url={https://arxiv.org/abs/2403.02310}, 
}

@misc{deep_speed_inference,
      title={DeepSpeed Inference: Enabling Efficient Inference of Transformer Models at Unprecedented Scale}, 
      author={Reza Yazdani Aminabadi and Samyam Rajbhandari and Minjia Zhang and Ammar Ahmad Awan and Cheng Li and Du Li and Elton Zheng and Jeff Rasley and Shaden Smith and Olatunji Ruwase and Yuxiong He},
      year={2022},
      eprint={2207.00032},
      archivePrefix={arXiv},
      primaryClass={cs.LG},
      url={https://arxiv.org/abs/2207.00032}, 
}

@inproceedings{xue2025regate-powergating-overhead,
title={ReGate: Enabling Power Gating in Neural Processing Units},
author={Xue, Yuqi and Huang, Jian},
booktitle={Proceedings of the 58th IEEE/ACM International Symposium on Microarchitecture},
pages={1160--1177},
year={2025}
}
